\documentclass[apj]{emulateapj}
\usepackage{mathtools}
\usepackage{natbib}
\usepackage{hyperref}
\usepackage{color}

\newcommand\invisiblesection[1]{%
\addcontentsline{toc}{section}{\protect\numberline{\thesection}#1}
\sectionmark{#1}}

\invisiblesection{foobar}

\shorttitle{The Red-Sequence in Galaxy Clusters}
\shortauthors{Foltz, Rettura et al.}

\begin{document}

\title{Evidence for the Universality of Properties of Red-Sequence Galaxies in X-ray- and Red-Sequence-Selected Clusters at z $\sim 1$}

\author{R. Foltz}
\affil{Department of Physics and Astronomy, University of California Riverside, 900 University Avenue, Riverside, CA 92521}
\email{ryan.foltz@email.ucr.edu}

\author{A. Rettura}
\affil{Infrared Processing and Analysis Center, California Institute of Technology, KS 314-6, Pasadena, CA 91125, USA}
\email{arettura@astro.caltech.edu}

\author{G. Wilson}
\affil{Department of Physics and Astronomy, University of California Riverside, 900 University Avenue, Riverside, CA 92521}
\email{gillian.wilson@ucr.edu}

\author{R.F.J. van der Burg}
\affil{
Laboratoire AIM, IRFU/Service d'Astrophysique - CEA/DSM - CNRS - Universit\'e Paris Diderot, B\^at. 709, CEA-Saclay, 91191 Gif-sur-Yvette Cedex, France
}
\email{remco.van-der-burg@cea.fr}

\author{A. Muzzin}
\affil{Institute of Astronomy, University of Cambridge, Madingley Road, Cambridge, CB3 0HA, United Kingdom}
\email{avmuzzin@ast.cam.ac.uk}

\author{C. Lidman}
\affil{Australian Astronomical Observatory, PO Box 915, North Ryde NSW 1670, Australia}
\email{clidman@aao.gov.au}

\author{R. Demarco}
\affil{Department of Astronomy, Universidad de Concepcion, Barrio Universitario. Casilla 160-C, Concepcion, Chile}
\email{rdemarco@astro-udec.cl}

\author{Julie Nantais}
\affil{
Grupo Astronom\'{i}a, Departamento de Ciencias F\'{i}sicas, Universidad Andr\'{e}s Bello, Rep\'{u}blica 220, Santiago, Chile}
\email{julie.nantais@unab.cl}

\author{A. DeGroot}
\affil{Department of Physics and Astronomy, University of California Riverside, 900 University Avenue, Riverside, CA 92521}
\email{adegr001@ucr.edu}

\author{H. Yee}
\affil{Dept of Astronomy and Astrophysics, University of Toronto, 50 Saint George Street, Toronto, ON M5S 3H4, Canada}
\email{hyee@astro.utoronto.ca}

\begin{abstract}
We study the slope, intercept, and scatter of the color-magnitude and color-mass relations for a sample of ten infrared red-sequence-selected clusters at z $\sim 1$. The quiescent galaxies in these clusters formed the bulk of their stars above $z \gtrsim 3$ with an age spread $\Delta t \gtrsim 1$ Gyr.
We compare UVJ color-color and spectroscopic-based galaxy selection techniques, and find a ~15\% difference in the galaxy populations classified as quiescent by these methods.
We compare the color-magnitude relations from our red-sequence selected sample with X-ray- and photometric-redshift-selected cluster samples of similar mass and redshift.
Within uncertainties, we are unable to detect any difference in the ages and star formation histories of quiescent cluster members in clusters selected by different methods,
suggesting that the dominant quenching mechanism is insensitive to cluster baryon partitioning at z$\sim 1$.
\end{abstract}


\section{Introduction}
Galaxy clusters form from the gravitational collapse and clustering of fluctuations in the primordial density field \citep{Press:1974aa,Gott:1975aa,Kravtsov:2012aa}.
This purely gravitational process is accompanied by the interrelated evolution of cluster baryonic components, including gas-dynamical radiative cooling and dissipation \citep{White:1978aa,Rudd:2008aa,Gnedin:2004aa,Kravtsov:2012aa}, and the formation and accretion of stellar mass in the form of galaxies. The resulting mature galaxy clusters are massive dark matter halos with deep gravitational potential wells, containing a hot intracluster medium (ICM), old, evolved galaxies, and intergalactic stars.

Surveys detect clusters via their baryonic components: originally clusters were identified by visual overdensity of galaxies in the optical regime \citep{Gunn:1986zm,Abell:1989lm,Lidman:1996uh,Ostrander:1998fx,Gal:2000lo}, and later by detection of the X-ray luminosity generated by the ICM \citep{Gioia:1994jm,Scharf:1997ws,Rosati:1998om,Romer:2000vp}. More recently, surveys have been designed to exploit the universal presence of a population of massive, quiescent galaxies (known as the red-sequence) in clusters \citep{Gladders:2000rq,Gladders:2005dp,Wilson:2009ws,Muzzin:2009jm,Rykoff:2014aa}, while others make use of overdensities of photometric redshifts \citep{Stanford:2005xr,Eisenhardt:2008pi}, or Sunyaev-Zeldovich (SZ) upscattering of cosmic microwave background photons by the ICM \citep{Reichardt:2013zn,Hasselfield:2013bs,Planck-Collaboration:2014aa,Bleem:2015sf}.

Because these cluster detection methods select on baryons rather than halo mass, each is inherently biased toward either gas-rich systems (as with X-ray or SZ methods) or galaxy-rich systems (for red-sequence and photometric-redshift methods).
Differences between gas-selected and galaxy-selected clusters are readily apparent, for example, when comparing the X-ray luminosity of the ICM, or cluster richness \citep{Donahue:2001fv,2008ApJ...675.1106R,Hicks:2013zl}.

As a galaxy falls into a cluster, interactions with cluster baryons can rapidly shut off its star formation in a process known as ``quenching''. This environmental quenching is driven by either the cluster galaxies (e.g. harassment \citep{Moore:1996aa}, tidal stripping \citep{Merritt:1983aa}, and mergers \citep{Toomre:1972aa,Rudnick:2012aa}) or the hot gas component (e.g. ram pressure stripping \citep{Gunn:1972aa,Quilis:2000aa}). One might expect that quenched galaxies in galaxy-rich red-sequence- or photometric-redshift-selected clusters would differ in their properties (e.g. luminosity-weighted ages and rest-frame colors) from quenched galaxies in gas-rich X-ray- or SZ-selected clusters, due to differences in quenching mechanisms and efficiencies.

Any such difference in the quiescent cluster galaxy populations will be more apparent at high redshift, when the galaxies are younger.
While X-ray cluster surveys have detected clusters out to $z \sim 1$, the launch of the Infrared Array Camera \citep[IRAC, ][]{Fazio:2004aa} on board the \textit{Spitzer Space Telescope} \citep{Werner:2004aa} has allowed systematic infrared red-sequence surveys to detect clusters at $z > 1$. {\it Spitzer}/IRAC wide-area surveys have proven effective at identifying more clusters down to low masses at $1  < z < 2$
\citep[e.g.,][]{Papovich:2010yj,Stanford:2012yi,Zeimann:2012bf,muzzin2013,Wylezalek:2013aa,Rettura:2014tt}, where current X-ray and SZ observations are restricted to only the most massive systems. We now have the opportunity to study quenched galaxies in red-sequence- and X-ray-selected clusters at $z = 1$, spanning the extremes of cluster baryon partitioning.

The focus of our study is the Gemini Cluster Astrophysics Spectroscopic Survey \citep[GCLASS \footnote{http://www.faculty.ucr.edu/\~{}gillianw/GCLASS/}, PIs: Wilson \& Yee][]{Muzzin:2012dw}. GCLASS is a sample of 10 red-sequence-selected clusters at $0.87 < z < 1.34$, initially detected by the SpARCS optical/IR cluster survey using the cluster red-sequence detection method developed by \cite{Gladders:2000rq} \citep[see][]{Muzzin:2009jm,Wilson:2009ws,Demarco:2010om}.
Our comparison sample is drawn from the ACS Intermediate Redshift Cluster Survey (ACS IRCS) \citep{fordACS}, a sample of six X-ray-selected and two optically-selected clusters at $0.8 < z < 1.3$, spanning a comparable range of redshifts and cluster masses. Five of the clusters were identified from the ROSAT Deep Cluster Survey \citep{Rosati:1998om}, while MS 1054-03 comes from the Einstein Extended Medium Sensitivity Survey \citep{Gioia:1994rz} and the clusters CL 1604+4304 and CL 1604+4321 were found in a Palomar deep near-infrared photographic survey \citep{Gunn:1986zm}. Extensive spectroscopic follow-up campaigns were conducted to assign cluster membership in these systems \citep[e.g.,][]{Demarco:2005aa,Demarco:2007aa,Holden:2006aa,Gal:2008aa,Rettura:2010aa}.

The red-sequence of galaxies is defined by a relation between the color and magnitude of quiescent galaxies \citep{Bower:1992mb,van-Dokkum:1998wd,Baldry:2004oq,Bell:2004qe}, and the slope and intrinsic scatter of this relation has been used to constrain the formation epochs and age spread of the early-type populations within galaxy clusters \citep{Bower:1998cr,2003ApJ...596L.143B,Mei:2009wt}. We will extend this study to our sample of red-sequence-selected clusters, comparing the formation redshifts and age spread constraints derived from our sample against those found for the comparison sample and others. If the red-sequence method preferentially selects older, more evolved systems, we would expect the red-sequence in these clusters to appear redder and exhibit less intrinsic scatter than other clusters at similar redshift.

The structure of this paper is as follows:
Our data set is described in Section \ref{data}.
In Section \ref{analysis}, we describe our cataloging and our methods for deriving galaxy rest-frame colors and stellar masses, as well as our fitting of the color-magnitude and color-mass relations. We compare spectroscopic and rest-frame UVJ selection methods in section \ref{sec-uvj}.
The results of our study are discussed in Section \ref{comparison}, while in Section \ref{conclusions} we summarize our conclusions.

In this work we will assume a standard $\Lambda$CDM cosmology with $H_0 = 70 \mathrm{\ km \cdot s^{-1} \cdot Mpc^{-1}}, \Omega_M = 0.3, \mathrm{\ and\ } \Omega_\Lambda = 0.7$, and a Chabrier IMF \citep{chabrierimf} throughout. Our magnitudes are reported in the AB system, unless reported otherwise.

\invisiblesection{Data}
\addtocounter{section}{1}
\subsection{Data}\label{data}
\begin{figure*}
\centering \includegraphics[width=\textwidth]{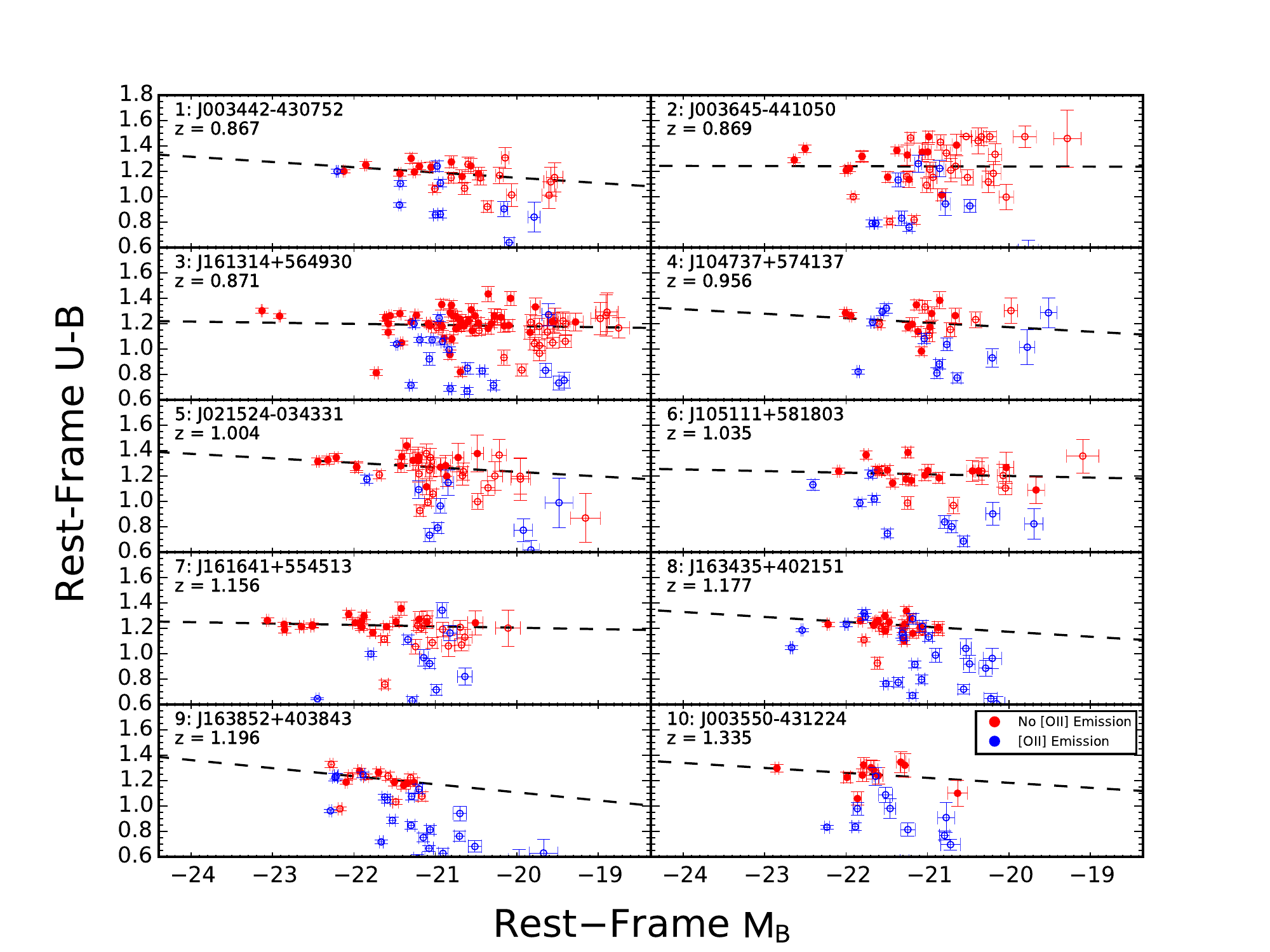}
\caption{Rest-frame U-B color versus absolute rest-frame B magnitude for spectroscopic members of each of the ten clusters in the GCLASS sample.
Quiescent members are shown in red. Those within R$_{200}$ above the 80\% mass completeness limit (see Table \ref{tbl-gclass}) are shown as solid. [O{\sc ii}]-emitters are shown in blue. The dashed lines show the Bayesian maximum likelihood linear fits to the color-magnitude relation for quiescent galaxies within R$_{200}$ above the 80\% mass completeness limit (solid red circles). See also Section \ref{sec-fits} and Table \ref{tbl-colormag}.
Note that some galaxies are classified as active because they have [O{\sc ii}] emission lines despite having colors consistent with red-sequence quiescent members. These could be be AGN, dust-obscured star-forming galaxies, or red-sequence objects with some residual star formation. They are not included in the fit.\label{fig-colormag}}
\end{figure*}

The red-sequence-selected clusters studied in this work are taken from GCLASS. GCLASS is a spectroscopic survey of 10 rich clusters at 0.85 $<$ z $<$ 1.34 which were initially detected by the SpARCS optical/IR cluster survey using the cluster red-sequence detection method developed by \cite{Gladders:2000rq} \citep[see][]{Muzzin:2009jm,Wilson:2009ws,Demarco:2010om}.

This cluster sample was the focus of a large spectroscopic campaign performed with the Gemini North and South Observatories. A total of 46 masks were observed over the ten clusters with the goal of identifying $\sim$ 50 members in each cluster. Galaxies were selected for spectroscopic follow-up through a combination of three criteria: distance from the cluster center, observed z' - 3.6$\mu$m color, and 3.6$\mu$m flux. Together these criteria ensure that the spectroscopic completeness is largely a function of stellar mass and radius, with the highest completeness found for massive galaxies in the cluster core \citep{Muzzin:2012dw}. 

The spectroscopic confirmation of these clusters was followed by optical imaging in $u^\prime\ g^\prime\ r^\prime\ i^\prime$ bands. For the six northern clusters, these data were taken with MegaCam at the Canada-France-Hawaii Telescope (CFHT), while for the southern clusters these data come from IMACS at Magellan/Baade ($u^\prime\ g^\prime\ r^\prime\ i^\prime$) . WIRCam at CFHT provided near-infrared \textit{J}- and \textit{Ks}-band data for the northern clusters, while for the southern clusters these data came from HAWK-I at the Very Large Telescope (VLT) or ISPI at CTIO/Blanco \citep{van-der-Burg:2013zn}. Our photometry also includes the 3.6, 4.5, 5.8, and 8.0 $\mathrm{\mu}$m IRAC data from the Spitzer Wide-area Extragalactic Survey (SWIRE, \citet{Lonsdale:2003ow}) and $z^\prime$ band data from the SpARCS survey taken by the MOSAIC-II camera at the Cerro Tololo Inter-American Observatory (CTIO)
\citep[see][for details on the Northern and Southern $z^\prime$ observations, respectively]{Muzzin:2009jm,Wilson:2009ws}.

A summary of the GCLASS sample is presented in Table \ref{tbl-gclass}. For more details on the spectroscopic and photometric observations we refer the reader to \citet{Muzzin:2012dw} and \citet{van-der-Burg:2013zn}, respectively.  The GCLASS sample has been used to study brightest cluster galaxies \citep[][Rettura et al., in prep.]{Lidman:2012il,Lidman:2013vp}, the relative effect of environment quenching and stellar mass quenching on galaxy evolution \citep{Muzzin:2012dw}, cluster and field stellar mass functions at $z\sim1$ \citep{van-der-Burg:2013zn}, cluster scaling relationships \citep{van-der-Burg:2014bs}, and phase space analysis constraints on the locations and timescales of quenching \citep{Noble:2013vh,Muzzin:2014sf}.

\subsection{Analysis}\label{analysis}

\subsubsection{Photometric Catalog}\label{sec-catalog}

As described in detail in \citet{van-der-Burg:2013zn}, the imaging data were combined into a matched catalog by first using Source Extractor to detect sources in the \textit{K$_s$}-band data. Each object was assigned a Gaussian PSF weight function based upon the Source Extractor half-light radius in the Ks-band image. This function was used to compute the weighted Gaussian-aperture-and-PSF flux for each object \citep{Kuijken:2008vf}. Robust errors were calculated by directly measuring the 1-$\sigma$ variation in background flux in randomly-placed apertures that do not contain any sources.

Galaxies are considered to be cluster members if they have a velocity relative to the cluster of $\Delta$ v $\le 1500$ km s$^{-1}$. For each cluster, this velocity dispersion was measured using the bi-weight estimator \citep{Beers:1990bf}
from the line-of-sight velocity distributions, after rejecting outliers \citep{Girardi:1993ud,Fadda:1996kk}. R$_{200}$, the radius for which the mean density is 200 times the critical density at the cluster redshift, was calculated assuming spherical clusters and the \citet{Evrard:2008lr} relation between $\mathrm{M}_{200}$ and the velocity dispersion (see Wilson et al., 2015, in prep. for details). Altogether this yields a total of 432 spectroscopically-confirmed cluster members for the full sample of ten clusters.


The photometric data were then matched to the spectroscopic catalog, such that each spectroscopic member considered in this work also has associated photometry in \textit{u' g' r' i' z' J Ks } and 3.6, 4.5, 5.8, 8.0 $\mu$m. In the following sections, we employ this catalog to derive rest-frame colors and magnitudes, stellar masses, and to fit the color-magnitude and color-mass relations for our cluster sample.

\subsubsection{Rest-Frame Colors}\label{sec-rest_frame_color}

In order to compare the photometric properties of cluster members over the range $0.8 < z < 1.3$, we need to first derive absolute rest-frame colors and magnitudes for these galaxies, to eliminate the effects of distance and account for k-correction.

To derive rest-frame photometry, we use the publicly-available photometric redshift code EAZY \citep{Brammer:2008uk}, to fit the broadband photometry of each cluster member to a linear combination of seven basis templates derived from the prescription in \citet{Blanton:2007ft}. These templates have been optimized for deep optical-NIR broad-band surveys, and this code was optimized specifically for K-selected samples such as our own. Comparing against the SEDs of galaxies in the GOODS-CDFS field, \citet{Brammer:2008uk} have shown that this method provides rest-frame optical photometry that is accurate to within 5\%.

\begin{figure*}
\centering \includegraphics[width=\textwidth]{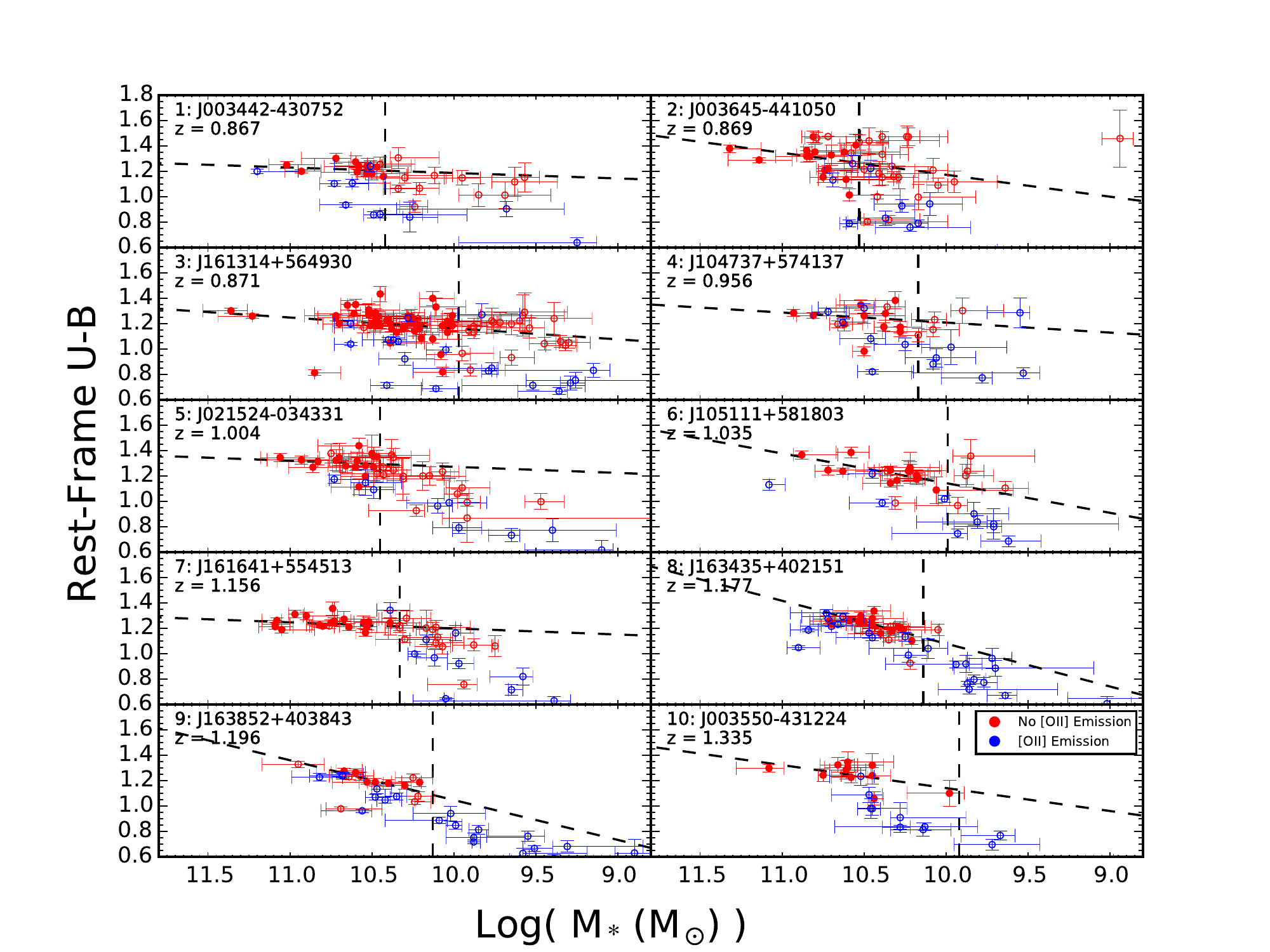}
\caption{Color-mass relation for each cluster in the GCLASS sample. Symbols are as in Figure 1. The dashed lines show the Bayesian maximum likelihood linear fits to the color-mass relation for quiescent galaxies within R$_{200}$ above the 80\% mass completeness limit (solid red circles). See also Section \ref{sec-fits} and Table \ref{tbl-colormass}.
Note that some galaxies are classified as active because they have [O{\sc ii}] emission lines despite having colors consistent with red-sequence quiescent members. These could be be AGN, dust-obscured star-forming galaxies, or red-sequence objects with some residual star formation. They are not included in the fit.\label{fig-colormass}}
\end{figure*}

We infer rest-frame absolute Johnson U and B magnitudes ($M_{U,z=0}$ , $M_{B,z=0}$) for each cluster member in our sample by convolving this best-fit linear combination of SEDs with filter curves redshifted to the spectroscopic redshift of each galaxy. These filters are chosen in order to directly compare results with those from the X-ray selected ACS IRCS sample \citep[][hereafter M09]{Mei:2009wt}. We define the rest frame color $\mathrm{(U-B)}_{z=0}$ as the difference between these rest-frame magnitudes. We note that the span of our 11 observed filters ensures that rest-frame magnitudes are interpolated from the available data, often overlapping with multiple observed passbands.

The rest frame $M_{B,z=0}$ magnitudes and $\mathrm{(U-B)}_{z=0}$ colors of cluster members are plotted in Figure \ref{fig-colormag}. Quiescent members are shown in red. Those within R$_{200}$ above the 80\% mass completeness limit (see Table \ref{tbl-gclass}) are shown as solid. [O{\sc ii}]-emitters are shown in blue. Some objects, while having colors consistent with red-sequence quiescent members, are nevertheless classified as active due to the presence of [O{\sc ii}] emission lines; these might be AGN, dust-obscured star-forming galaxies, or red-sequence objects with some residual star formation. They are not included in the CMR fit.

The estimated 1-$\sigma$ uncertainties in these rest-frame colors and magnitudes are derived from 200 Monte Carlo simulations: the observed flux in each passband is varied by a random amount drawn from a normal distribution with a standard deviation given by the photometric uncertainties, and the CMR fit repeated.
The central 68\% of the resulting output then defines the upper and lower confidence intervals on our rest-frame $(U-B)_{z=0}$ color.

\subsubsection{Stellar masses}\label{sec-stellar_masses}

Here we use our photometric catalogs to derive stellar masses for the cluster members for the purpose of fitting the color-mass relations.

Using the publicly-available SED fitting code FAST \citep{Kriek:2009eq}, we fit the 11-passband photometry to \citet{Bruzual:2003ge} (BC03) stellar populations synthesis templates. Although other stellar models employ a different treatment for contributions from thermally pulsing asymptotic giant branch (TP-AGB) phase stars, there is disagreement about the significance of TP-AGB stars for galactic SEDs and inferred galactic properties \citep{Kriek:2010fq}. The population models with a different treatment of the TP-AGB phase, such as those from \citet{Maraston:2005oj}, yield stellar masses that are lower by 0.1 dex on average. These values are consistent within error bars with those derived from BC03, and the choice of model does not significantly impact our results at the redshift range considered in this study \citep[see e.g.][]{Rettura:2006gb}.

FAST proceeds by generating a grid of synthetic SEDs for stellar populations at the spectroscopic redshift of each galaxy from the given population synthesis templates, for a range of star formation histories (SFH), ages, and masses, with possible additional variation in dust attenuation and/or metallicity. Best-fit stellar populations are then selected from this grid by minimizing ${\chi}^2$ when comparing the SED to the observed broad-band photometry of a given galaxy, providing us with an estimate of stellar mass.

For our grid of parameters, we use a range of ages from 10 Myr to 10 Gyr (excluding ages greater than the age of the universe at the observed redshift) and an A$_V$ ranging from 0 to 4 mag with a Calzetti extinction law \citep{Calzetti:2001hh}. An exponentially declining star formation rate is assumed with a time constant, $\tau$, ranging from 10 Myr to 10 Gyr. A Chabrier IMF \citep{chabrierimf} and fixed (solar) metallicity of 0.02 is assumed throughout.

With the best fit stellar mass thus derived, a confidence interval is provided by a Monte Carlo simulation with 200 iterations. The color-mass relations are plotted in Figure \ref{fig-colormass}.

\subsubsection{Color-Magnitude and Color-Mass Relations}\label{sec-fits}

Having explained how we derive rest-frame colors and magnitudes in Section \ref{sec-rest_frame_color}, and stellar masses in Section \ref{sec-stellar_masses}, we are now ready to determine the color-magnitude and color-mass relations.

From the spectroscopically-confirmed cluster members in each cluster, we select red-sequence galaxies for the purpose of fitting the CMR. A member galaxy is included in the fit if it meets the following three criteria:

\begin{enumerate}

\item
Quiescent galaxies are selected as those galaxies without any detected [O{\sc ii}] emission line, where the detection limit is $\sim -1 \mathrm{\AA}$ to $-3 \mathrm{\AA}$ equivalent width, depending on signal-to-noise \citep[see][]{Muzzin:2012dw}.

\item
We select galaxies above the $80\%$ mass completeness limit for each cluster \citep[calculated in][]{van-der-Burg:2013zn}.

\item
For each cluster, we fit to quiescent galaxies within R$_{200}$. The GCLASS clusters, despite their richness, display a variety of morphologies and so the centroid of the cluster was taken to be the brightest cluster galaxy \citep{Lidman:2012il}. The R$_{200}$ radii for the GCLASS sample were taken from Wilson et al. (2015, in prep).
\end{enumerate}

For each cluster, quiescent galaxies selected as described above (solid red circles in Figure \ref{fig-colormag}) were fit by a rest-frame $(U-B)_{z=0}$ versus $M_{B,z=0}$ color-magnitude relation of the following form:

\begin{equation}\label{eq-colormag}
(U-B)_{z=0} = \textrm{slope} \times (M_{B,z=0} + 21.4) + c_0
\end{equation}

where $M_{B,z=0}$ is the rest-frame B magnitude, and $c_0$ is the CMR zeropoint. A magnitude offset of 21.4 is applied to reduce the covariance of the slope and zeropoint. The specific value of 21.4 is taken from M09, to allow for direct comparison.

It has been noted for some time that the choice of linear regression method can bias the estimate of slopes and correlation coefficients \citep{Kelly:2007sf}. Where possible, we prefer to use a Bayesian maximum likelihood estimator (MLE) to yield the least biased values, and do so to arrive at the fits shown in Figures \ref{fig-colormag} and \ref{fig-colormass} and the analysis performed in Sections \ref{sec-CMR-age} and \ref{sec-scatevo}. However, in order to directly compare with M09, \citet{Bower:1992mb}, \citet{van-Dokkum:1998wd}, and \citet{Ellis:1997lk} (Section \ref{sec-evo}), we will use a total least squares (TLS) method to most closely match the comparison analysis.

Our TLS method derives uncertainties in the fit parameters using a bootstrapping method with 1000 simulations to calculate the 68\% confidence interval in slopes, scatters, and zeropoints. The intrinsic scatter of the relation is calculated by subtracting in quadrature the photometric error from the biweight scale estimate of the color residuals.

For the MLE method, we calculate the posterior probability distributions of the slope, zeropoint, and intrinsic scatter with a Markov chain Monte Carlo method. These probability distributions directly yield the most likely value of each fit parameter and the associated 68\% confidence interval. The MLE method has been shown to yield more unbiased fits and confidence intervals than other linear regression methods \citep{Kelly:2007sf}.

The regression parameters reported by each method are compatible within error bars, although the uncertainty of the MLE-derived CMR parameters increases by as much as $\sim 35 - 45\%$, as expected. The TLS method finds a slope that is $\sim 20\%$ steeper on average, and smaller intrinsic scatters throughout, while there is agreement to within $\sim 2\%$ for the average zeropoints.  The slopes, scatters, and zeropoints of the CMRs derived for each of the GCLASS clusters using both methods are reported in Table \ref{tbl-colormag}.

The properties of the CMR depend in part on the morphology of the galaxies that are included in the red-sequence. A red-sequence including both lenticular and elliptical galaxies (E+S0) typically exhibits $\sim 30\%$ more scatter than a purely elliptical sample  \citep[see e.g.][M09]{Bower:1992mb}. In all cases, our results will be compared to literature values derived for E+S0 galaxies, and we discuss this comparison in Section \ref{sec-evo}.

In addition to making a fit to the color-magnitude relation for each cluster (Figure \ref{fig-colormag}), we also make a fit to the color-mass relation (Figure \ref{fig-colormass}). We employ a fit of the form:

\begin{equation}\label{eq-colormass}
(U-B)_{z=0} = \mathrm{slope} \times (\mathrm{Log}(M_*/M_{\odot})-10.6) + c_0
\end{equation}

where we have chosen the mass offset of 10.6 dex so that the average color-mass zeropoint corresponds to the average color-magnitude zeropoint, described in greater detail in Section \ref{sec-colormass}.

We employ a Bayesian maximum likelihood estimator to obtain the fit parameters. As we do not compare our color-mass fits to literature values, this single fitting method will be sufficient for our purposes and yield the least biased results.

We show the color-mass relations in Figure \ref{fig-colormass}. The fit parameters are reported in Table \ref{tbl-colormass}, and discussed in Section \ref{sec-colormass}.

\subsubsection{Rest-frame UVJ Diagrams}\label{sec-uvj}
\begin{figure*}
\centering \includegraphics[width=\textwidth]{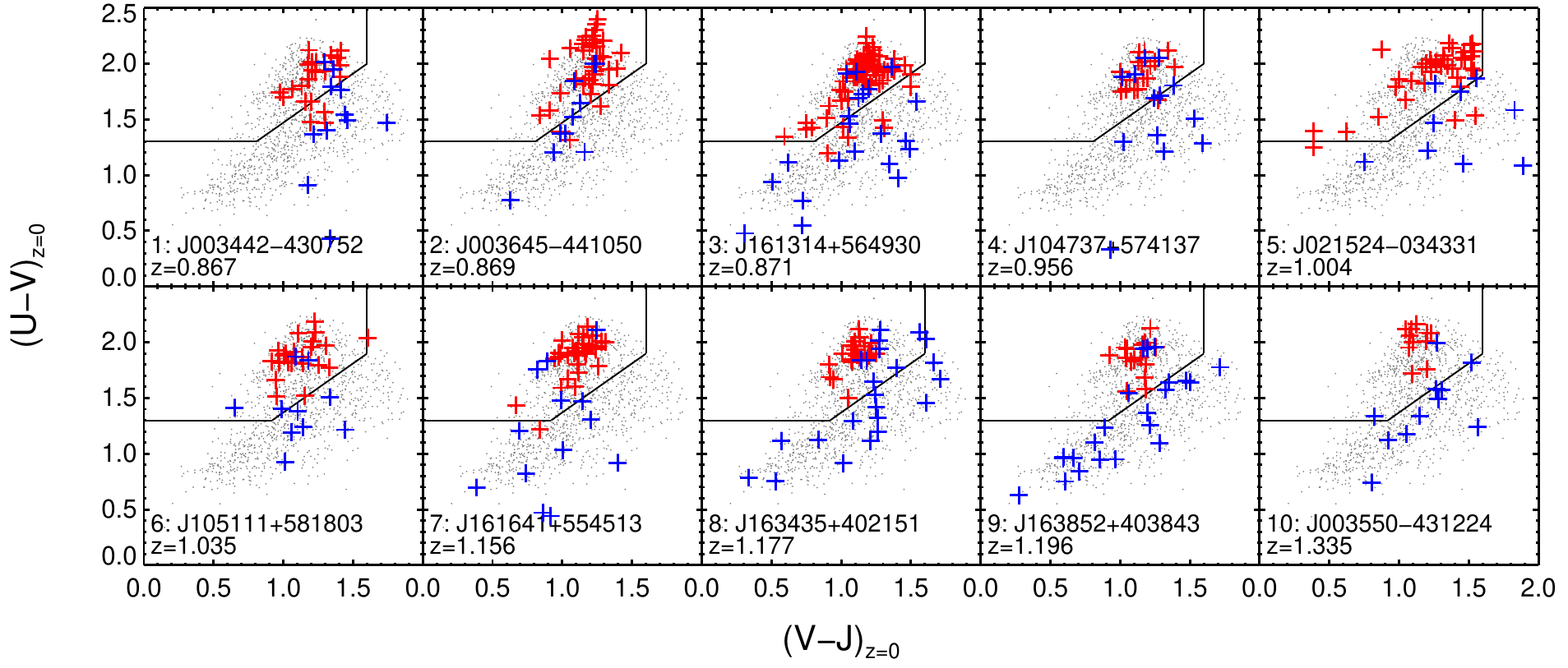}
\caption{Rest-frame U-V versus V-J color for each cluster in the GCLASS sample. Quiescent cluster members (without [O{\sc ii}] emission lines) are plotted as red crosses, while star-forming cluster members are plotted in blue. The solid lines show the color-color cut used by \citet{Williams:2009tt} for distinguishing quiescent from star-forming galaxies, where the upper-left quadrant is typically populated by quiescent galaxies.
We plot in gray a sample of field galaxies from GCLASS with masses $M_* > 10^{9.5} M_\odot$ that are between $0.85 < z < 1.35$, to illustrate the color space occupation at the redshift of the clusters.
14\% of the UVJ-quiescent population show [O{\sc ii}] emission, while 16\% of the UVJ-star-forming galaxies exhibit no [O{\sc ii}] emission.
\label{fig-uvj}}
\end{figure*}

\begin{figure*}
\centering \includegraphics[width=\textwidth]{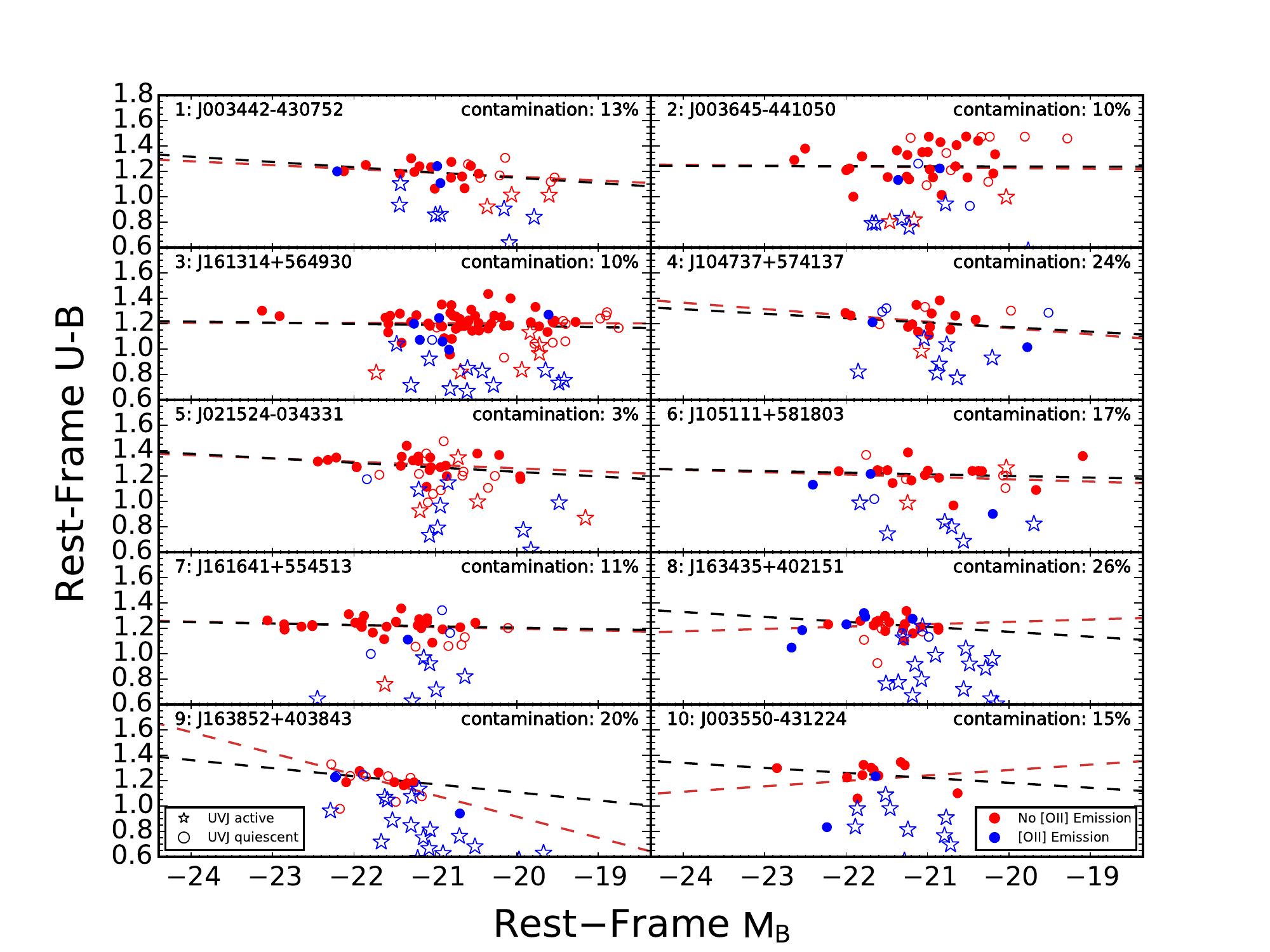}
\caption{
Figure showing how red sequence slope would differ if quiescent members are selected on UVJ color rather than spectroscopically (compare red line to black line from Figure \ref{fig-colormag}).
As in Figure \ref{fig-colormag}, the figure shows rest-frame U-B versus rest-frame absolute B magnitude for spectroscopic members, color coded by spectroscopically quiescent (red) and star forming (blue), but with symbols denoting UVJ quiescent (circles) and active (stars).  The indicated contamination measures the proportion of UVJ-quiescent, spectroscopically-active galaxies (blue circles).
The fit shown as a dashed red line is to the solid circles which are UVJ-quiescent galaxies within R$_{200}$ above the 80\% mass completeness limit. See discussion in Section \ref{sec-uvj}.
\label{fig-uvjcolormag}}
\end{figure*}

Rest-frame UVJ color-color selection is often used to distinguish quiescent and star-forming galaxies in field galaxy surveys at redshifts $0 < z < 4$ when spectroscopic or morphological information is not available \citep{Wuyts:2007aa,Williams:2009tt,Whitaker:2011aa,Patel:2012ab,van-der-Burg:2013zn,Whitaker:2013rz,muzzin2013,Strazzullo:2013aa}.
Because the GCLASS dataset consists of a large number of both quiescent and active spectroscopically-confirmed members, the GCLASS dataset allows a unique insight into the efficacy of the UVJ selection technique, by allowing us to compare the degree of agreement between UVJ- and spectrally-classified active / quiescent members.

We plot all of the spectroscopically-confirmed cluster members in rest-frame UVJ color-color space in Figure \ref{fig-uvj}.
As in Figures \ref{fig-colormag} and \ref{fig-colormass}, members are colored according to spectroscopic classification, with red being quiescent and blue showing [O{\sc ii}] emission lines (see Section \ref{sec-fits}).

Those galaxies falling in the upper-left quadrant of this plot are classified as UVJ-quiescent according to the cuts described in \citet{Williams:2009tt}. There is a clear, but not exact, agreement between the spectroscopic and UVJ methods: 14\% of the UVJ-quiescent population show [O{\sc ii}] emission, while 16\% of the UVJ-star-forming galaxies exhibit no [O{\sc ii}] emission. Although we do not expect much contamination from AGN, an [O{\sc ii}] selection nevertheless excludes these objects. The main source of contamination for an [O{\sc ii}] selection is likely dusty star-forming galaxies. While the UVJ selection accounts for dust, uncertainty in rest-frame colors will naturally result in some cross-contamination between the quiescent and star-forming populations, especially for galaxies close to the dividing lines. For our sample, a spectroscopic [O{\sc ii}]-based selection is more stringent than one based on UVJ colors, ultimately identifying fewer quiescent galaxies.

This level of contamination is similar to that found by other studies \citep{kriek2014}. In a large sample of galaxies spanning a redshift range $0.8 \leq z \leq 1.2$, \citet{Cardamone:2010aa} finds that 20\% of galaxies on the red-sequence fall outside the UVJ-quiescent selection region. \citet{Moresco:2013aa} study the relative agreement of different quiescent selection methods, finding [O{\sc ii}] emission lines for 38\% of the UVJ-quiescent galaxies, and an overall $20-40\%$ contamination for this method. Additionally, from a total sample of of 19 cluster galaxies at $z = 1.80$, \citet{Newman:2014aa} find $12\%$ of members classified as UVJ-star-forming when they are spectroscopically quiescent.

To better understand the implications of UVJ versus spectroscopic selection methods, we repeat the CMR relationship fitting of Section \ref{sec-fits}, now using the UVJ classification instead of the [O{\sc ii}] spectral feature to select quiescent galaxies. The resulting fits are shown in Figure \ref{fig-uvjcolormag}.

In this figure, as in Figures \ref{fig-colormag}, \ref{fig-colormass}, and \ref{fig-uvj}, members with an [O{\sc ii}]-emission line are shown in blue while spectroscopically-quiescent members are shown in red. At the same time, we represent UVJ-quiescent members as circles and UVJ-active members as stars. Therefore, UVJ-selected quiescent galaxies that were previously excluded by the [O{\sc ii}]-selection of Section \ref{sec-fits} are shown as blue circles. The relative number of these objects is labeled as the quiescent contamination for each cluster. Lastly, the UVJ-quiescent galaxies that satisfy the remaining two criteria of Section \ref{sec-fits}, being within R$_{200}$ and above the 80\% mass completeness limit, are solid symbols. The MLE CMR fit to these solid circles is shown as a dashed red line, contrasted by the black line which shows the fits derived previously in Section \ref{sec-fits} for the [O{\sc ii}]-selected quiescent galaxies.

The intrinsic scatter is comparable for both selections, although the zeropoint and slope of the UVJ-selected red-sequences are more discrepant over our range of clusters, exhibiting larger uncertainties and a wider spread in values. For most clusters, the small number of contaminants introduced by UVJ selection has little impact on the CMR slope.
The largest discrepancies are seen in our highest-redshift clusters which have fewer spectroscopic members overall, where the inclusion or removal of one or two galaxies can dramatically impact the resulting CMR fit (Figure \ref{fig-uvjcolormag}).

We conclude that while the rest-frame UVJ selection technique is generally effective in separating quiescent from star-forming galaxies in the absence of spectroscopy, we caution that users should expect a non-negligible amount of contamination ($\sim 15\%$ in this case).


\section{Discussion} \label{comparison}

\subsection{Comparison of Color-Magnitude Relation as Function of Cluster Selection Method}\label{sec-evo}

\begin{figure}
\includegraphics[width=0.5\textwidth]{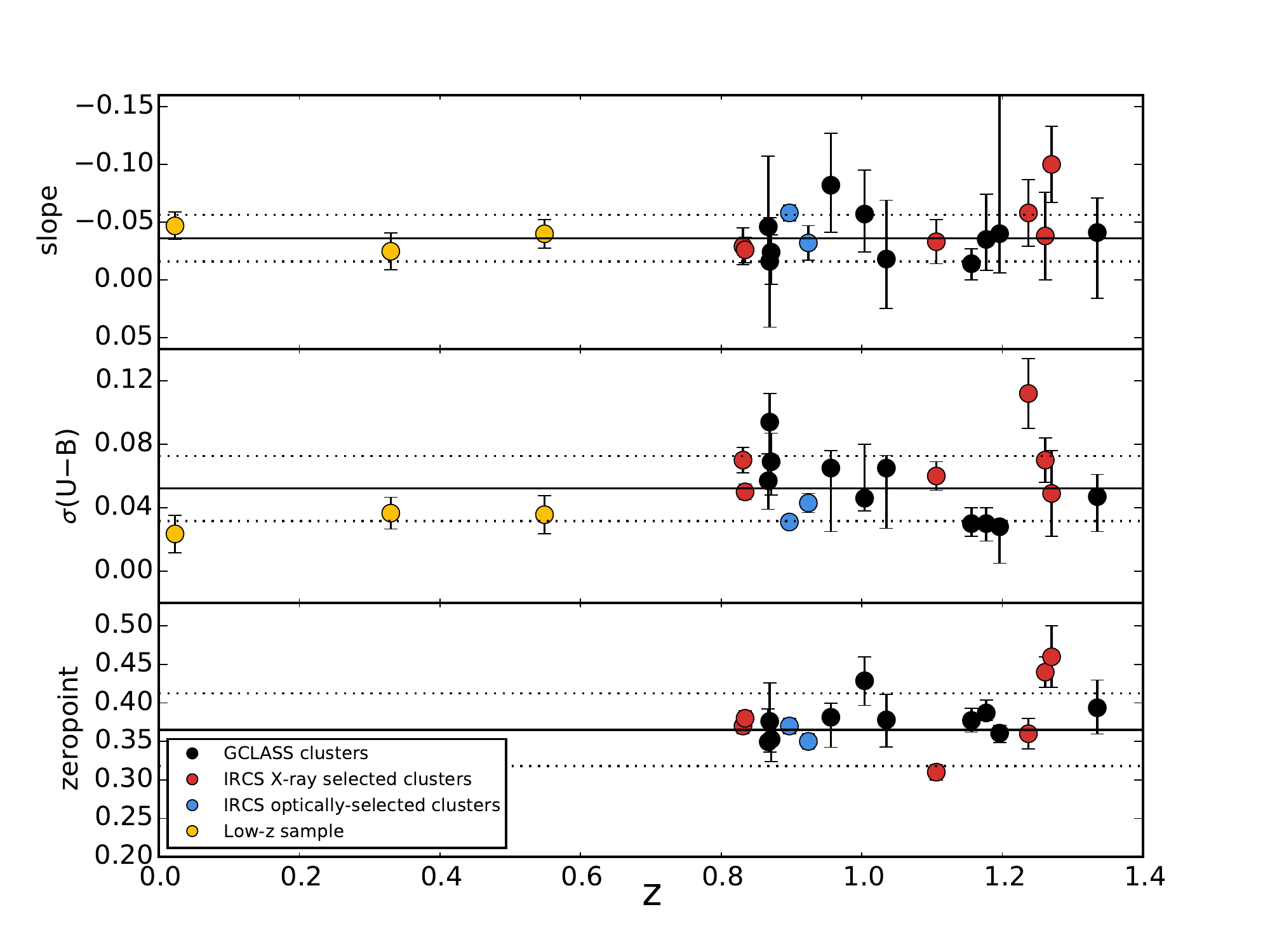}
\caption{Evolution of the color-magnitude relation slope (top), scatter (middle), and zeropoint (bottom) with redshift, compared to values drawn from the literature. Data points from the GCLASS clusters are plotted as black circles, data from the ACS Intermediate Redshift Survey as red circles (X-ray-selected) or blue circles (optically-selected) (M09), and data from a local sample \citep{Bower:1992mb,van-Dokkum:1998wd,Ellis:1997lk} as yellow circles. The biweight mean value of the parameters from the ACS IRCS sample of X-ray-detected clusters is plotted as a solid line, with dotted lines indicating the 1-$\sigma$ range. We find no measurable difference in the slope, zeropoint or scatter of the CMR in clusters selected by the X-ray or red-sequence technique (see Section \ref{sec-evo}) The zeropoints shown here are Vega magnitudes.
\label{fig-params}}
\end{figure}

Figure \ref{fig-params} shows the red-sequence slope, intrinsic scatter, and zeropoint for the ten GCLASS clusters plotted versus the cluster redshift. Also shown are the data from the ACS IRCS sample \citep[][M09]{fordACS}, including six X-ray selected and two optically-selected clusters spanning a redshift range $0.83 < z < 1.3$. We also plot data from the local clusters Coma, Virgo, and CL0016 (z = 0.546) \citep{Bower:1992mb,van-Dokkum:1998wd,Ellis:1997lk}. For comparison, we plot the biweight mean and 1-$\sigma$ variation of the ACS IRCS sample parameters. Within uncertainties, the data points for all GCLASS clusters fall within the 1-$\sigma$ variation of the biweight mean values of the ACS IRCS sample.

We find no significant difference between the red-sequences found in X-ray- and red-sequence-selected galaxy clusters. A Student's $t$-test reveals agreement between the fit parameters we report versus those drawn from the ACS IRCS sample ($t=0.89, p=38\%$). We conclude that the stellar populations of quiescent galaxies in these clusters selected by different methods have comparable histories of stellar formation and evolution.
We can detect no indication that these quiescent populations have experienced different quenching histories or processes, and it is likely that the dominant quenching process does not depend on the baryon partitioning of the cluster.
If differences in the quenching mechanism or history are present, they are not detectable in the resultant red-sequence properties at $z \sim 1$, at least at the resolution of our observations.

Overall, the X-ray and red-sequence methods are selecting clusters with relatively similar CMRs and small intrinsic scatters, which indicates a high formation redshift for the bulk of the massive galaxy populations in these clusters.
Perhaps surprisingly, the agreement in CMR zeropoint between these samples indicates that the red-sequence method is not selecting clusters with older, more evolved populations.
While the methods may still select different clusters by mass, evolutionary state, or virialization, those properties do not seem to correlate with the stellar populations of the galaxies.

In addition, we find that the CMR slope and scatter have not significantly evolved since redshift $z\sim1.3$. This is in agreement with prior studies of clusters for z $< 1.5$ \citep[e.g. M09, ][]{Snyder:2012wq}.
For these parameters, a Bayesian model comparison substantially favors a constant over a model that is linear with redshift. Over the redshift range we sample, we find no measurable evolution in the zeropoint ($0.8 < z < 1.3$).

\subsection{Red-Sequence Ages}\label{sec-CMR-age}

The $(U-B)_{z=0}$ color of the galaxies we have been studying is the result of several competing influences: principally, it will be determined by the galaxy metallicity in a mass-dependent fashion; it will also redden as young, blue stars die out or transition onto the red giant branch, or become bluer as ongoing star formation provides new, massive stars. The redshift evolution of the CMR (in slope, zeropoint, and scatter) is therefore sensitive to all of these factors, and we can partially constrain the formation and evolution of early-type galaxies by comparing our results to models.

Our models have two parameters: their formation redshift and star formation history (SFH). For the choice of SFH, we generate models with both single-burst simple stellar populations (SSP) and exponentially-declining star formation histories (eSFH). The formation redshifts range from $ 2 \leq z_f \leq 9$. For details on the construction of these models, see Appendix \ref{appendix}.

The zeropoint of the CMR models is simply the $(U-B)_{z=0}$ color evaluated at $M_{B}=-21.4$. The zeropoint is a measure of the average color of the red-sequence and therefore is sensitive to both the age of the galaxies and the extent of their most recent star formation. For this reason it can constrain the star formation weighted age, $\langle t \rangle_\mathrm{SFH}$, which gives the average age of the bulk of the stars. From this age we can also establish a star formation weighted formation redshift, $\langle z_f \rangle_\mathrm{SFH}$. See Appendix \ref{appendix} for the definition of these quantities.

Altogether, within the error bars of our measurement, we find no significant evolution of the zeropoint with redshift for our clusters. As $(U-B)_{z=0}$ color evolves most rapidly for young stellar populations, this implies that the bulk of the stars must have formed at sufficiently high redshift so that their color is only slowly evolving by $z \sim 1$. We note that this is a constraint on the overall age of the stars, while the galaxies were likely assembled some time after the initial star formation.

Although the eSFH models redden more slowly, both SSP and eSFH models agree on a lower-bound for $\langle z_f \rangle_\mathrm{SFH} \sim 3$ for our highest redshift clusters. Essentially, this means that if the red-sequence galaxies were still forming stars below $z \sim 3$, the high-redshift CMRs would appear bluer than what we observe in GCLASS.

\begin{figure}
\includegraphics[width=0.5\textwidth]{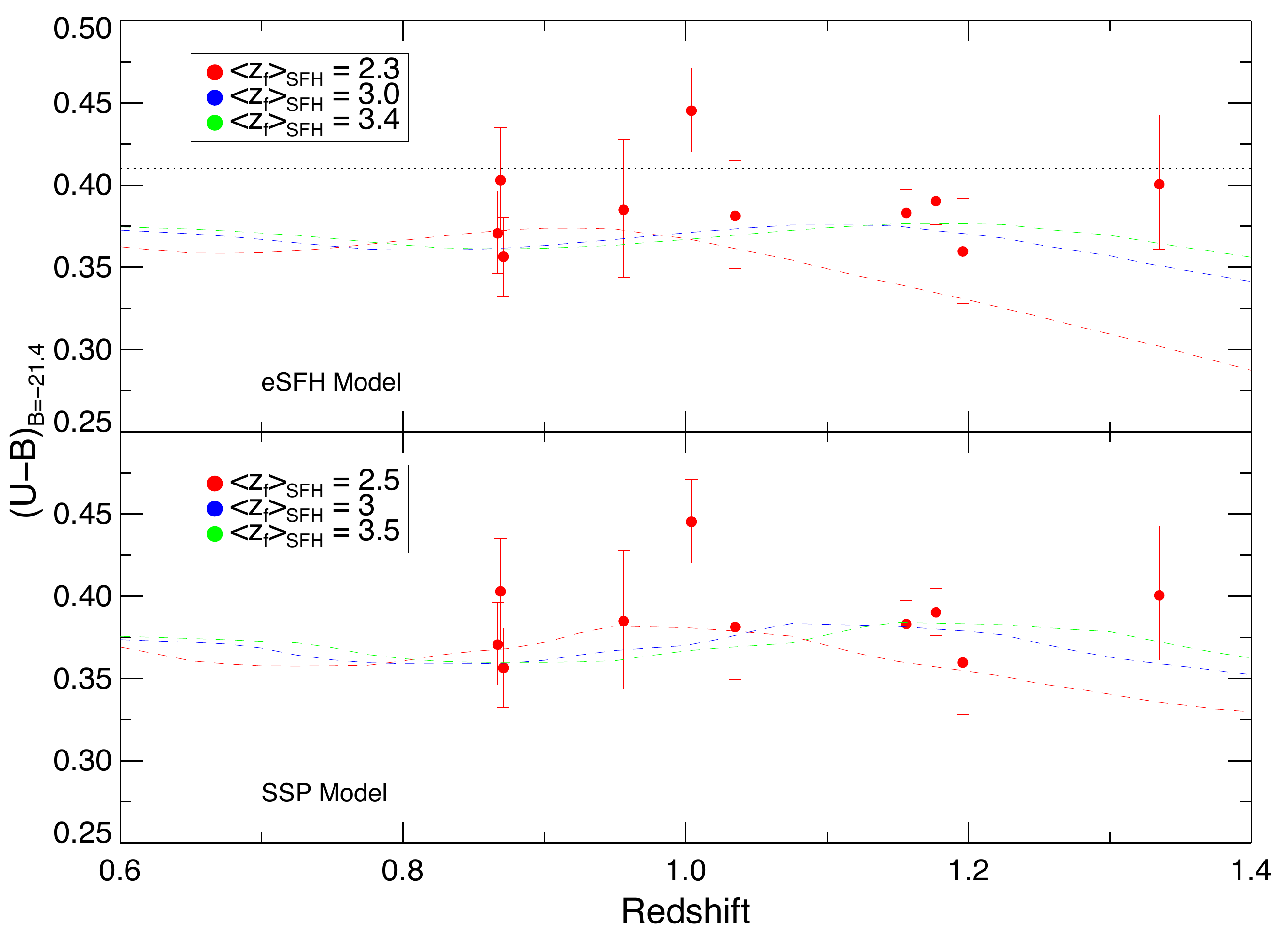}
\caption{Evolution of the color-magnitude relation zeropoint with redshift, compared with exponential SFH (top) and SSP (bottom) model red-sequences. Measured zeropoints of the GCLASS clusters are plotted as red circles in the same manner as the bottom panel of figure \ref{fig-params}. Model values are plotted as dashed lines, while the biweight mean and 1-$\sigma$ variation of our zeropoint values are plotted as solid and dotted lines, respectively. The minimal value for $\langle z_f \rangle_\mathrm{SFH}$ that agrees with the data is $\sim 3$.
\label{fig-models}}
\end{figure}

\begin{figure}
\includegraphics[width=0.5\textwidth]{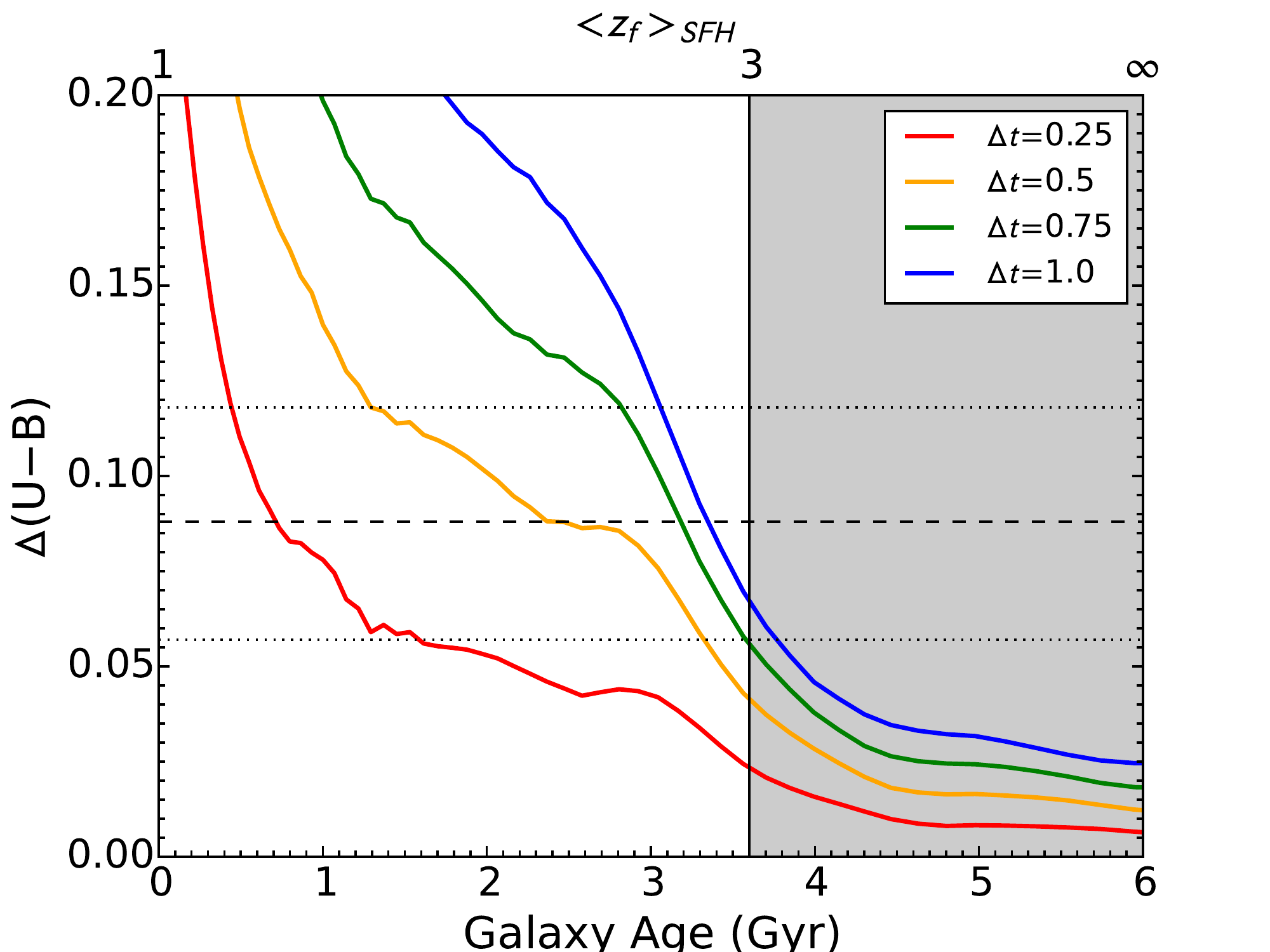}
\caption{Difference in $(U-B)_{z=0}$ color for pairs of model stellar populations with age differences $\Delta t = 0.25, 0.5, 0.75$, and $1.0$ Gyr. The age of the younger galaxy model is plotted on the x axis. The vertical solid line is placed at the lower-limit age constraint derived from our zeropoint considerations, such that all viable models must lie in the shaded region. The horizontal lines indicate the biweight mean and 1-$\sigma$ variation in our MLE-measured $(U-B)_{z=0}$ scatters. From this plot, we see that our measured intrinsic scatter is consistent with an age spread $\Delta t > 1$ Gyr.
\label{fig-scatmodel}}
\end{figure}

If we assume that the principle cause of the intrinsic scatter about the CMR is due to a spread in galaxy ages, it is then possible to constrain this spread by comparing our observed intrinsic scatter with red-sequence models. To do this, we first quantify the difference in $(U-B)_{z=0}$ color between two galaxies, $\Delta(U-B)_{z=0}$, as a function of time and of the difference in the galaxies' ages, $\Delta t$, and relate this to the observed scatter.

Since $(U-B)_{z=0}$ color evolves most quickly for young stellar populations, the largest $\Delta(U-B)_{z=0}$ color differences are apparent when both galaxies are young. However, a larger difference in galaxy ages also yields a higher $\Delta(U-B)_{z=0}$ overall, leading to a degeneracy in the color difference / age difference relation which can be partially broken by using our previous zeropoint constraint to require that our galaxies be at least as old as $\langle t \rangle_\mathrm{SFH}$. At $z \sim 1$, our lower bound of $\langle z_f \rangle_\mathrm{SFH} \sim 3$ corresponds to a minimum galaxy age of 3.6 Gyr. The age of the universe at $z \sim 1$ provides an upper bound on galaxy age differences to be below $\Delta t \lesssim 2 $ Gyr at the most extreme.

The time evolution of $\Delta(U-B)_{z=0}$ color differences for pairs of galaxies of different ages is plotted in Figure \ref{fig-scatmodel}.

The evolution of this color scatter, and our determination of an age spread, depends on our choice of star formation history. Generally, models with larger $\tau$ evolve more slowly, allowing our average $\Delta(U-B)_{z=0}$ color scatter to correspond to a larger spread in age. However, taking all models into account, and assuming that the intrinsic scatter is an age effect, we can constrain the average age spread to be $\Delta t \gtrapprox 1$ Gyr. Note that this is a lower bound : if we allow our galaxies to be older than our lower bound limit of $\langle t \rangle_\mathrm{SFH} = 3.6$ Gyr,  we could recover the same $\Delta(U-B)_{z=0}$ color scatter by simply allowing a larger spread in ages. 

\subsection{Scatter Evolution}\label{sec-scatevo}

\begin{figure}
\includegraphics[width=0.5\textwidth]{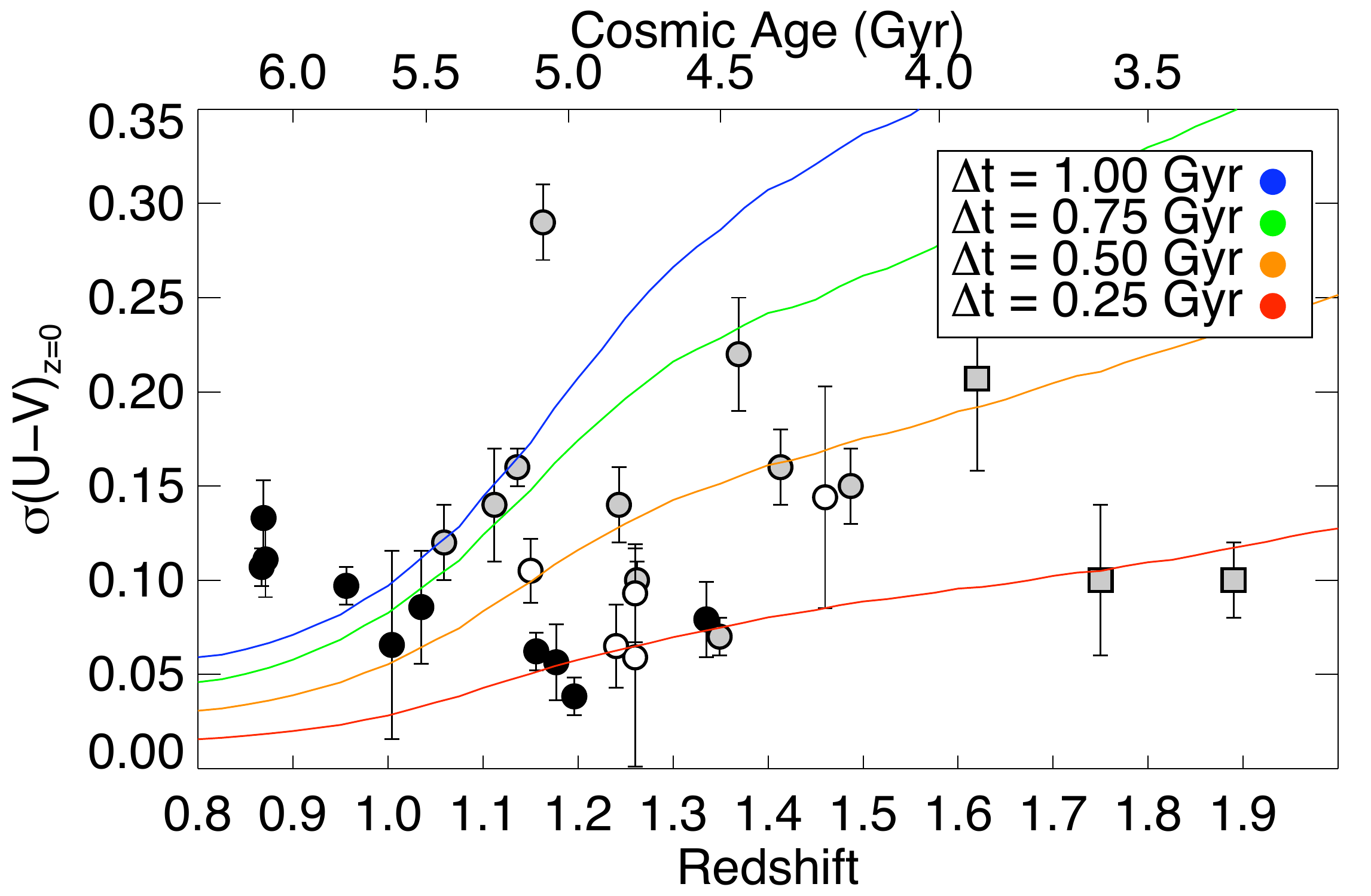}
\caption{Evolution of the color-magnitude relation $(U-V)_{z=0}$ scatter with redshift, compared against values drawn from \citet{Snyder:2012wq} (gray circles), \citet{Hilton:2009nq} (white circles), and \citet{Papovich:2010yj}, \citet{Stanford:2012yi}, \citet{Zeimann:2012bf} (gray squares). Data from the GCLASS sample are plotted as black circles. The overlaid models show the difference in $(U-V)_{z=0}$ color for pairs of galaxies of different ages ($\Delta t$), and therefore trace the passive evolution of a purely age-dependent scatter. The models are averaged over galaxies for $\langle z_f \rangle_\mathrm{SFH} > 2$.
\label{fig-scatter_evolution}}
\end{figure}

It is clear from Figure \ref{fig-scatmodel} that color differences due to galaxy age differences are larger when galaxies are younger, and for this reason we expect the intrinsic scatter of the CMR to increase with redshift. However, this is at odds with the apparent lack of evolution in scatter with redshift over our sample (Figure \ref{fig-params}).

In Figure \ref{fig-scatter_evolution} we plot the intrinsic scatter of our sample against that measured by \citet{Snyder:2012wq} and the data compiled by \citet{Hilton:2009nq}, including clusters from \citet{Mei:2006ec} and \citet{Blakeslee:2003qq}, and high-redshift clusters from \citet{Papovich:2010yj}, \citet{Stanford:2012yi}, \citet{Zeimann:2012bf}. The scatter data are reported for the $(U-V)_{z=0}$ CMR, so we repeat our rest-frame color derivation and CMR fitting using Johnson U and V filters.

In general, the rest-frame colors and intrinsic scatters for these clusters are not determined in a homogeneous manner. Where we have estimated the rest-frame colors from 11-band SED-fitting, rest-frame colors frequently are determined by a linear conversion of an observed color using a synthetic color model. Intrinsic scatter has historically been measured in a variety of different ways, sometimes employing reduced chi-square normalization, although the technique used in this work is more common today. Various criteria are also used to select red-sequence galaxies, either photometrically or morphologically, with different completeness limits. The use of color or $\sigma$ cuts can bias the measurement of intrinsic scatter.

The overlaid models represent the simple evolution of the $\Delta (U-V)_{z=0}$ color difference between pairs of galaxies of different ages ($\Delta t$). The color differences of these model galaxy pairs can be interpreted as evolution tracks of the CMR intrinsic scatter for passively-evolving red-sequences, if we allow that the color scatter is a measure of the red-sequence age spread. While these predicted scatters are dependent to some degree on the formation redshift, the effect is small relative to the inherent uncertainty involved in calculating the intrinsic scatter of the CMR, and so we average the color evolution models over formation redshifts for $\langle z_f \rangle_\mathrm{SFH} > 2$.

The models show an increase in scatter with redshift in a manner that depends on $\Delta t$. As a whole, the reported scatter values broadly exhibit an increasing trend, in agreement with \citet{Hilton:2009nq}, possibly indicating the expected passive evolution. Clearly, no single evolutionary track can connect all of the galaxy clusters in the GCLASS sample, nor is this larger literature sample consistent with a single history. 

A possible explanation for this is that our sample exhibits progenitor bias \citep{2000ApJ...541...95V,2001ASPC..230..581F} and that younger, bluer galaxies are migrating onto the red-sequence as they are quenched. This would naturally result in an age spread that increases with redshift, which is consistent with the trend seen in Figure \ref{fig-scatter_evolution}. Furthermore, clusters continuously accrete field galaxies, which may have metallicities and ages that are different from the galaxies within the cluster. The introduction of these galaxies would also potentially increase the scatter.

\subsection{The Color-Mass Relation}\label{sec-colormass}

\begin{figure}[t]
\includegraphics[width=0.5\textwidth]{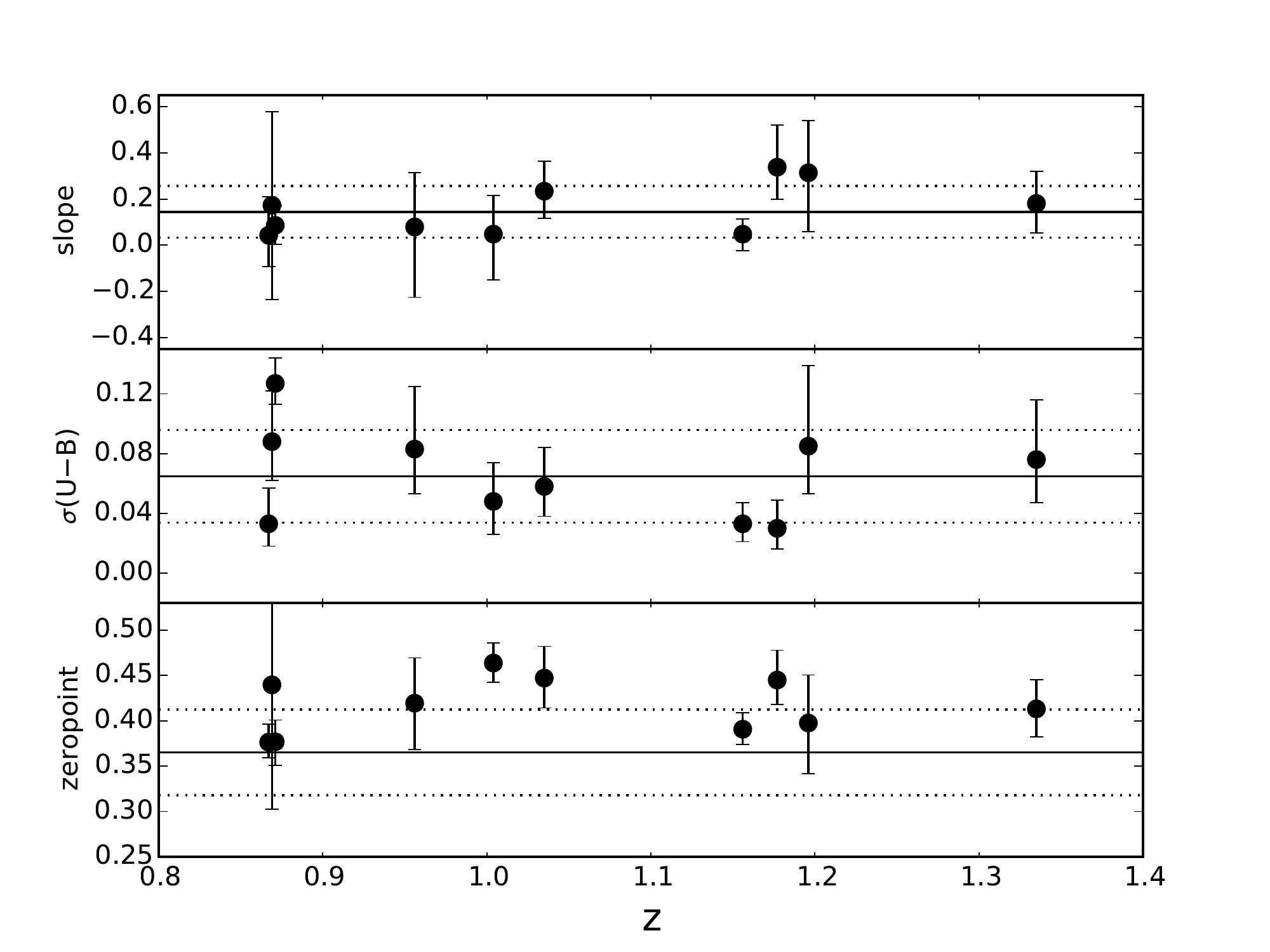}
\caption{Evolution of the color-mass relation slope (top), scatter (middle), and zeropoint (bottom) with redshift. The biweight mean value of the parameters is plotted as a solid line, with dotted lines indicating the 1-$\sigma$ range.\label{fig-massparams}}
\end{figure}

Historically, studies have compared galaxy color as a function of magnitude, with models to infer the properties of quiescent members. However, galaxy properties such as morphology, size, and color also correlate with stellar mass, and it is a more physically meaningful parameter than magnitude. In Figure \ref{fig-colormass} we show the color-mass relation and linear fits for the ten GCLASS clusters (see Table \ref{tbl-colormass}). While previous studies have presented color-mass relations \citep{Borch:2006aa,Cardamone:2010aa,Huertas-Company:2010aa,Strazzullo:2010kh,Bassett:2013hq,Cimatti:2013aa,Moresco:2013aa}, this is the first time to our knowledge that the red-sequence color-mass relation has been fit.

The redshift evolution of this relation's slope, scatter, and zeropoint is shown in Figure \ref{fig-massparams}. We plot the biweight mean values of our fit parameters and their 1-$\sigma$ variation (see Table \ref{tbl-colormass}). We have been unable to find any comparable mass-color fits in the literature to compare with at low redshift, so we are only able to investigate the redshift evolution of the fit over the redshift range spanned by the GCLASS sample. Over this redshift range, $0.8 < z < 1.3$, we do not detect any evolution in the color-mass relation. We note that we also do not find any evolution in the color-mag relation over this redshift range (see Section \ref{sec-evo}).



\section{Conclusions}\label{conclusions}

In this paper we studied the rest-frame U-B color-magnitude and color-mass relations for 10 red-sequence-selected clusters between redshifts $0.8 < z < 1.3$.
We compared these results with those from the X-ray-selected ACS IRCS sample \citep{Mei:2009wt}, to look for differences in the properties of red-sequence galaxies in galaxy-selected and gas-selected clusters at $z \sim 1$.
From the data analysis presented in this work, we find the following conclusions:

\begin{itemize}

\item
The fact that we observe no measurable differences between the ages and CMR properties of quiescent cluster members in clusters selected by different methods suggests that, at least at z $\sim 1$, the dominant quenching mechanism is insensitive to cluster baryon partitioning, favoring processes such as preprocessing or strangulation.
The remarkable agreement in color-magnitude zeropoint throughout these cluster samples indicates that the red-sequence method does not preferentially select older, more evolved systems.

\item
The CMR zeropoints measured for $0.8 < z < 1.3$ allow us to constrain the quiescent members' period of last major star formation to be above $z \sim 3$ for our high-redshift clusters. The observed intrinsic scatters of the CMR in our cluster sample are indicative of an average age spread greater than 1 Gyr.

\item The lack of evolution in the intrinsic scatter over $0.8 < z < 1.3$ cannot be explained by simple passive evolution of the red-sequence, indicating possibly a progenitor bias created by younger galaxies migrating onto the red-sequence. This process would result in intrinsic scatters that are consistent with larger age spreads at the lower-redshift end of the sample.

\item UVJ color-color classification of quiescent and star-forming galaxy populations broadly agrees with spectroscopic classification based on [O{\sc ii}] emission. From a total sample of 432 spectroscopically-confirmed cluster members, 14\% of the UVJ-quiescent population show [O{\sc ii}] emission, while 16\% of the UVJ-star-forming galaxies exhibit no [O{\sc ii}] emission.

\item We present the color-mass relationship and linear fit parameters for the GCLASS sample. We detect no measurable evolution of the color-mass relationship over the redshift range of the sample, $0.8 < z < 1.3$. The intrinsic scatter of the color-mass relationship agrees with that measured for the color-magnitude relation.

\end{itemize}

\acknowledgments{
Financial support for this work was provided by NSF grants AST-0909198 and AST-1517863.

Support for this work was provided by an award issued by JPL/Caltech. RvdB acknowledges support from the European Research Council under FP7 grant number 340519. R.D. gratefully acknowledges the support provided by the BASAL Center for Astrophysics and Associated Technologies (CATA), and by FONDECYT grant N. 1130528.
J.N. acknowledges support from FONDECYT postdoctoral research grant N. 3120233.

We thank Simona Mei for providing the exact filter definition files used in her previous work.

Based on observations obtained with MegaPrime / MegaCam, a joint project of CFHT and CEA/IRFU, at the Canada-France-Hawaii Telescope (CFHT) which is operated by the National Research Council (NRC) of Canada, the Institut National des Science de l'Univers of the Centre National de la Recherche Scientifique (CNRS) of France, and the University of Hawaii. This work is based in part on data products produced at Terapix available at the Canadian Astronomy Data Centre as part of the Canada-France-Hawaii Telescope Legacy Survey, a collaborative project of NRC and CNRS.
Based on observations obtained with WIRCam, a joint project of CFHT, Taiwan, Korea, Canada, France, at the Canada-France-Hawaii Telescope (CFHT) which is operated by the National Research Council (NRC) of Canada, the Institut National des Sciences de l'Univers of the Centre National de la Recherche Scientifique of France, and the University of Hawaii. This work is based on observations obtained at the CTIO Blanco 4-m telescopes, which are operated by the Association of Universities for Research in Astronomy Inc. (AURA), under a cooperative agreement with the NSF as part of the National Optical Astronomy Observatories (NOAO). This paper includes data gathered with the 6.5 meter Magellan Telescopes located at Las Campanas Observatory, Chile. Based on observations that were carried out using the Very Large Telescope at the ESO Paranal Observatory. This work is based in part on observations made with the Spitzer Space Telescope, which is operated by the Jet Propulsion Laboratory, California Institute of Technology under a contract with NASA. Support for this work was provided by NASA through an award issued by JPL/Caltech. Based on observations obtained at the Gemini Observatory, which is operated by the Association of Universities for Research in Astronomy, Inc., under a cooperative agreement with the NSF on behalf of the Gemini partnership: the National Science Foundation (United States), the National Research Council (Canada), CONICYT (Chile), the Australian Research Council (Australia), Ministério da Ciência, Tecnologia e Inovação (Brazil) and Ministerio de Ciencia, Tecnología e Innovación Productiva (Argentina).}

\appendix

\section{Star Formation Histories}\label{appendix}
The eSFH models employ a star formation history parametrized by:
\begin{equation}\label{eq-esfh}
\Psi(t,\tau) = \mathrm{SFR}_0\cdot e^{-t/\tau} \left[ \frac{M_{\odot}}{\mathrm{yr}} \right],
\end{equation}
where $\Psi$ is the instantaneous star formation rate at time $t$ after the onset of star formation, SFR$_0$ is the initial star formation rate, and $\tau$ is a parameter ranging from 0.5 Gyr $\leq \tau \leq$ 5 Gyr.

We begin by generating model stellar populations with the range of six metallicities provided by the BC03 population synthesis code ($\mathrm{Z} = 0.0001, 0.0004, 0.004, 0.008, 0.02\, (\mathrm{Z}_\odot) , 0.05$). These six models each exhibit a metallicity-dependent U-V, J-K, and V-K color, but the absolute $M_{B,z=0}$ magnitudes are free parameters of these stellar population models.
We therefore fix each $M_{B,z=0}$ magnitude by finding the value for which the reported Coma CMRs \citep{Bower:1992mb} best match each model's U-V, J-K, and V-K colors, by minimizing $\chi^2$.
This essentially provides mass normalizations for each galaxy model and reproduces the CMR at $z=0$. The galaxy models then describe a simple red-sequence which may be passively evolved backward in redshift to provide a predicted redshift evolution of CMR slope and zeropoint.

There is remarkably good agreement between the slope of the modeled red-sequence and those which we report for the GCLASS sample in Table \ref{tbl-colormag}. The modeled slope does not evolve significantly or disagree with our measured slope between $0.8 < z < 1.3$, for any chosen formation redshift or star formation history, and therefore does not constrain our models.

The competing influences of age and ongoing star formation on the zeropoint color introduce a sort of degeneracy which means we cannot distinguish between young galaxies that very quickly shut off star formation and old galaxies with more recent star formation. Therefore we combine both factors into a single parameter which can be constrained by observation: the star formation weighted age, $\langle t \rangle_\mathrm{SFH}$, which gives the average age of the bulk of the stars, following \citet{Rettura:2011aa}:

\begin{equation}\label{eq-tsfh}
\langle t \rangle_\mathrm{SFH} \equiv \frac{\int_0^t(t-t^\prime)\Psi(t^\prime,\tau)\mathrm{d}t^\prime}
{\int_0^t\Psi(t^\prime,\tau)\mathrm{d}t^\prime}
\end{equation}

For the star formation history defined in Equation \ref{eq-esfh}, this equals

\begin{equation}\label{eq-etsfh}
\langle t \rangle_\mathrm{SFH} = \frac{t - \tau + \tau\cdot e^{-t/\tau}}{1-e^{-t/\tau}} .
\end{equation}

Our observed zeropoints can then place constraints on this $\langle t \rangle_\mathrm{SFH}$, which, together with the lookback time to the redshift of the cluster, can constrain the star-formation-weighted formation redshift $\langle z_f \rangle_\mathrm{SFH}$ of these stellar populations. We note that in general, the formation of the stars will be followed by their assembly into galaxies, and the age of the galaxy will be younger than the age of its component stars.
\clearpage
\bibliography{mybib}

\clearpage

\begin{deluxetable}{llcccccc}
\tabletypesize{\scriptsize}
\tablecaption{The GCLASS cluster sample\label{tbl-gclass}}
\tablewidth{0pt}
\tablehead{
\colhead{Index}
& \colhead{Cluster}
& \colhead{R.A.}
& \colhead{Decl.}
& \colhead {z}
& \colhead{Members\textsuperscript{a}}
& \colhead{R$_{200}$\textsuperscript{b} ($h^{-1}$ Mpc) }
& \colhead{Mass ($10^{14} \mathrm{M}_\odot$)}
}
\startdata
1 & SpARCS J003442-430752 & 00:34:42.086 & -43:07:53.360 & 0.866 & 34 & $0.93\substack{+0.10\\-0.18}$ & $2.4\substack{+0.9\\-1.2}$ \\
2 & SpARCS J003645-441050 & 00:36:45.039 & -44:10:49.911 & 0.869 & 48 & $1.1\substack{+0.1\\-0.2}$    & $4.4\substack{+1.7\\-1.6}$ \\
3 & SpARCS J161314+564930 & 16:13:14.641 & 56:49:29.504  & 0.871 & 89 & $1.8\substack{+0.1\\-0.2}$    & $17.6\substack{+4.6\\-4.2}$ \\
4 & SpARCS J104737+574137 & 10:47:37.463 & 57:41:37.960  & 0.956 & 30 & $0.76\substack{+0.09\\-0.10}$ & $1.5\substack{+0.6\\-0.5}$ \\
5 & SpARCS J021524-034331 & 02:15:23.200 & -03:43:34.482 & 1.004 & 46 & $0.55\substack{+0.07\\-0.10}$ & $0.59\substack{+0.27\\-0.26}$ \\
6 & SpARCS J105111+581803 & 10:51:11.232 & 58:18:03.128  & 1.035 & 32 & $0.61\substack{+0.10\\-0.13}$ & $0.85\substack{+0.46\\-0.44}$ \\
7 & SpARCS J161641+554513 & 16:16:41.232 & 55:45:25.708  & 1.156 & 45 & $0.74\substack{+0.09\\-0.12}$ & $1.7\substack{+0.7\\-0.7}$ \\
8 & SpARCS J163435+402151 & 16:34:35.402 & 40:21:51.588  & 1.177 & 46 & $0.89\substack{+0.11\\-0.12}$  & $3.1\substack{+1.3\\-1.1}$ \\
9 & SpARCS J163852+403843 & 16:38:51.625 & 40:38:42.893  & 1.196 & 39 & $0.56\substack{+0.06\\-0.12}$ & $0.77\substack{+0.31\\-0.40}$ \\
10 & SpARCS J003550-431224 & 00:35:49.700 & -43:12:24.160 & 1.335 & 23 & $1.0\substack{+0.2\\-0.2}$   & $5.5\substack{+3.0\\-3.2}$ \\
\enddata
\tablenotetext{a}{Number of spectroscopically-confirmed member galaxies.}
\tablenotetext{b}{The radius for which the mean density is 200 times the critical density. From Wilson et al. (2015, in prep).}
\end{deluxetable}

\clearpage

\begin{deluxetable}{llcccccc}
\tabletypesize{\scriptsize}
\tablecaption{Color-Magnitude Fit Parameters\label{tbl-colormag}}
\tablewidth{0pt}
\tablehead{
\colhead{Index} & \colhead{Cluster}
& \colhead{N\textsuperscript{a}}
& \colhead{Method\textsuperscript{b}}
& \colhead{$\frac{ \Delta(U-B)_{z=0} }{ \Delta M_{B,z=0} }$\textsuperscript{c}}
& \colhead{$\sigma(U-B)_{z=0}$\textsuperscript{d}}
& \colhead{$c_0 \mathrm{(AB)}$\textsuperscript{e}}
& \colhead{$c_0 \mathrm{(Vega)}$}
}
\startdata
1  & SpARCS J003442-430752 & 14 & MLE & $-0.041\substack{+0.043 \\ -0.044} $ & $0.067\substack{+0.024 \\ -0.017}$ & $1.211\substack{+0.026 \\ -0.023}$ & $0.3720\substack{+0.026 \\ -0.023}$ \\[1.5ex]
   &                       &    & TLS & $-0.046\substack{+0.030 \\ -0.061} $ & $0.057\substack{+0.017 \\ -0.018}$ & $1.189\substack{+0.043 \\ -0.006}$ & $0.3490\substack{+0.043 \\ -0.006}$ \\[1.5ex]
2  & SpARCS J003645-441050 & 22 & MLE & $-0.003\substack{+0.041 \\ -0.042} $ & $0.123\substack{+0.028 \\ -0.022}$ & $1.239\substack{+0.031 \\ -0.028}$ & $0.4020\substack{+0.031 \\ -0.028}$ \\[1.5ex]
   &                       &    & TLS & $-0.016\substack{+0.057 \\ -0.038} $ & $0.094\substack{+0.018 \\ -0.024}$ & $1.214\substack{+0.050 \\ -0.040}$ & $0.3760\substack{+0.050 \\ -0.040}$ \\[1.5ex]
3  & SpARCS J161314+564930 & 54 & MLE & $-0.008\substack{+0.023 \\ -0.024} $ & $0.123\substack{+0.015 \\ -0.013}$ & $1.194\substack{+0.025 \\ -0.023}$ & $0.3570\substack{+0.025 \\ -0.023}$ \\[1.5ex]
   &                       &    & TLS & $-0.024\substack{+0.028 \\ -0.015} $ & $0.069\substack{+0.018 \\ -0.021}$ & $1.191\substack{+0.018 \\ -0.029}$ & $0.3530\substack{+0.018 \\ -0.029}$ \\[1.5ex]
4  & SpARCS J104737+574137 & 14 & MLE & $-0.034\substack{+0.074 \\ -0.077} $ & $0.112\substack{+0.036 \\ -0.025}$ & $1.226\substack{+0.041 \\ -0.042}$ & $0.3870\substack{+0.041 \\ -0.042}$ \\[1.5ex]
   &                       &    & TLS & $-0.082\substack{+0.041 \\ -0.045} $ & $0.065\substack{+0.011 \\ -0.040}$ & $1.223\substack{+0.018 \\ -0.039}$ & $0.3810\substack{+0.018 \\ -0.039}$ \\[1.5ex]
5  & SpARCS J021524-034331 & 24 & MLE & $-0.033\substack{+0.043 \\ -0.044} $ & $0.102\substack{+0.025 \\ -0.020}$ & $1.284\substack{+0.028 \\ -0.026}$ & $0.4450\substack{+0.028 \\ -0.026}$ \\[1.5ex]
   &                       &    & TLS & $-0.057\substack{+0.033 \\ -0.038} $ & $0.046\substack{+0.034 \\ -0.008}$ & $1.269\substack{+0.031 \\ -0.032}$ & $0.4290\substack{+0.031 \\ -0.032}$ \\[1.5ex]
6  & SpARCS J105111+581803 & 16 & MLE & $-0.013\substack{+0.044 \\ -0.041} $ & $0.089\substack{+0.032 \\ -0.024}$ & $1.220\substack{+0.032 \\ -0.033}$ & $0.3820\substack{+0.032 \\ -0.033}$ \\[1.5ex]
   &                       &    & TLS & $-0.018\substack{+0.043 \\ -0.051} $ & $0.065\substack{+0.008 \\ -0.038}$ & $1.216\substack{+0.033 \\ -0.035}$ & $0.3780\substack{+0.033 \\ -0.035}$ \\[1.5ex]
7  & SpARCS J161641+554513 & 25 & MLE & $-0.009\substack{+0.017 \\ -0.018} $ & $0.046\substack{+0.012 \\ -0.011}$ & $1.222\substack{+0.015 \\ -0.014}$ & $0.3840\substack{+0.015 \\ -0.014}$ \\[1.5ex]
   &                       &    & TLS & $-0.014\substack{+0.014 \\ -0.013} $ & $0.030\substack{+0.010 \\ -0.008}$ & $1.215\substack{+0.016 \\ -0.015}$ & $0.3770\substack{+0.016 \\ -0.015}$ \\[1.5ex]
8  & SpARCS J163435+402151 & 17 & MLE & $-0.038\substack{+0.044 \\ -0.044} $ & $0.051\substack{+0.018 \\ -0.013}$ & $1.229\substack{+0.014 \\ -0.015}$ & $0.3900\substack{+0.014 \\ -0.015}$ \\[1.5ex]
   &                       &    & TLS & $-0.035\substack{+0.027 \\ -0.039} $ & $0.030\substack{+0.010 \\ -0.011}$ & $1.226\substack{+0.017 \\ -0.010}$ & $0.3870\substack{+0.017 \\ -0.010}$ \\[1.5ex]
9  & SpARCS J163852+403843 & 7 & MLE &  $-0.064\substack{+0.088 \\ -0.100} $ & $0.061\substack{+0.046 \\ -0.023}$ & $1.200\substack{+0.032 \\ -0.031}$ & $0.3590\substack{+0.032 \\ -0.031}$ \\[1.5ex]
   &                       &   & TLS &  $-0.040\substack{+0.034 \\ -0.140} $ & $0.028\substack{+0.001 \\ -0.023}$ & $1.200\substack{+0.010 \\ -0.012}$ & $0.3610\substack{+0.010 \\ -0.012}$ \\[1.5ex]
10  & SpARCS J003550-431224 & 11 & MLE & $-0.030\substack{+0.068 \\ -0.068} $ & $0.094\substack{+0.044 \\ -0.032}$ & $1.242\substack{+0.042 \\ -0.040}$ & $0.4030\substack{+0.042 \\ -0.040}$ \\[1.5ex]
    &                       &    & TLS & $-0.041\substack{+0.057 \\ -0.030} $ & $0.047\substack{+0.014 \\ -0.022}$ & $1.233\substack{+0.036 \\ -0.034}$ & $0.3940\substack{+0.036 \\ -0.034}$ \\
\enddata
\tablenotetext{a}{The number of quiescent cluster member galaxies used in computing the fit.}
\tablenotetext{b}{The method used to derive fit parameters: maximum likelihood estimator (MLE) or total least squares (TLS).}
\tablenotetext{c}{The slope of the rest-frame U-B color-magnitude relation.}
\tablenotetext{d}{The intrinsic scatter of the rest-frame U-B color-magnitude relation.}
\tablenotetext{e}{The zeropoint, i.e., the U-B color of the color-magnitude relation evaluated at M$_B$=-21.4, reported as an AB magnitude.}
\end{deluxetable}

\clearpage

\begin{deluxetable}{llcccc}
\tabletypesize{\scriptsize}
\tablecaption{Color-Mass Fit Parameters\label{tbl-colormass}}
\tablewidth{0pt}

\tablehead{
\colhead{Index} & \colhead{Cluster}
& \colhead{N\textsuperscript{a}}
& \colhead{$\frac{ \Delta(U-B)_{z=0} }{ \Delta log(M_*/M_{\odot}) }$\textsuperscript{b}}
& \colhead{$\sigma(U-B)_{z=0}$\textsuperscript{c}}
& \colhead{$c_0$\textsuperscript{d}}
}
\startdata
1  & SpARCS J003442-430752 & 18 & $0.043\substack{+0.166 \\ -0.136} $ & $0.033\substack{+0.024 \\ -0.015}$ & $1.211\substack{+0.020 \\ -0.017}$ \\
2  & SpARCS J003645-441050 & 28 & $0.173\substack{+0.405 \\ -0.408} $ & $0.088\substack{+0.034 \\ -0.026}$ & $1.267\substack{+0.095 \\ -0.137}$ \\
3  & SpARCS J161314+564930 & 44 & $0.086\substack{+0.085 \\ -0.083} $ & $0.127\substack{+0.017 \\ -0.014}$ & $1.209\substack{+0.024 \\ -0.026}$ \\
4  & SpARCS J104737+574137 & 15 & $0.079\substack{+0.235 \\ -0.305} $ & $0.083\substack{+0.042 \\ -0.030}$ & $1.252\substack{+0.050 \\ -0.051}$ \\
5  & SpARCS J021524-034331 & 30 & $0.048\substack{+0.168 \\ -0.199} $ & $0.048\substack{+0.026 \\ -0.022}$ & $1.298\substack{+0.022 \\ -0.021}$ \\
6  & SpARCS J105111+581803 & 14 & $0.234\substack{+0.130 \\ -0.118} $ & $0.058\substack{+0.026 \\ -0.020}$ & $1.271\substack{+0.035 \\ -0.033}$ \\
7  & SpARCS J161641+554513 & 29 & $0.048\substack{+0.065 \\ -0.071} $ & $0.033\substack{+0.014 \\ -0.012}$ & $1.225\substack{+0.018 \\ -0.017}$ \\
8  & SpARCS J163435+402151 & 17 & $0.338\substack{+0.182 \\ -0.140} $ & $0.030\substack{+0.019 \\ -0.014}$ & $1.263\substack{+0.033 \\ -0.027}$ \\
9  & SpARCS J163852+403843 & 7 & $0.314\substack{+0.226 \\ -0.255} $ & $0.085\substack{+0.054 \\ -0.032}$ & $1.217\substack{+0.053 \\ -0.056}$ \\
10  & SpARCS J003550-431224 & 11 & $0.181\substack{+0.139 \\ -0.128} $ & $0.076\substack{+0.040 \\ -0.029}$ & $1.240\substack{+0.032 \\ -0.031}$ \\
\enddata
\tablecomments{These fit parameters were derived using a Bayesian maximum likelihood estimator.}
\tablenotetext{a}{The number of quiescent cluster member galaxies used in computing the fit.}
\tablenotetext{b}{The slope of the rest-frame U-B color-mass relation.}
\tablenotetext{c}{The intrinsic scatter of the rest-frame U-B color-mass relation.}
\tablenotetext{d}{The zeropoint of the rest-frame U-B color-mass relation, in AB magnitudes.}
\end{deluxetable}

\end{document}